\documentstyle[aas2pp4]{article}          %%%<< 2column, for Preprint 

\def\r{\hangindent=1pc  \noindent}
\def\kms{km s$^{-1}$} 
\def\deg{$^\circ$}  
\def\Deg{^\circ}

\begin{document} 

\title{Bipolar-Hyper-Shell Galactic Center Statrburst Model: 
Further Evidence from ROSAT Data and New Radio and X-ray Simulations}
\author{{Yoshiaki SOFUE}\\
{\it Institute of Astronomy, University of Tokyo, Mitaka, Tokyo 181-8588}\\
{\it E-mail: sofue@ioa.s.u-tokyo.ac.jp} }
\date{}
%\maketitle 

\begin{abstract} 
Using the all-sky ROSAT soft X-ray and 408-MHz radio continuum
data, we show that the
North Polar Spur and its western and southern counter-spurs
draw a giant dumbbell-shape necked at the galactic plane.
We interpret these features as due to a shock front originating from
a starburst 15 million years ago with a total energy of the 
order of $\sim 10^{56}$ ergs or $10^5$ type II supernovae. 
We simulate all-sky distributions of radio continuum and soft X-ray 
intensities based on the bipolar-hyper-shell galactic center starburst model.
The simulations can well reproduce the radio NPS and related spurs, as well as
radio spurs in the tangential directions of spiral arms.
Simulated X-ray maps in 0.25, 0.75 and 1.5 keV bands reproduce the
ROSAT X-ray NPS, its western and southern counter-spurs, and the 
absorption layer along the galactic plane.
We propose to use the ROSAT all-sky maps to probe the physics of gas in the
halo-intergalactic interface, and to directly date and measure 
the energy of a recent Galactic Center starburst.
\end{abstract}

\section{Introduction}

A bipolar hyper-shell model  has been proposed for the North Polar Spur (NPS)
(Sofue 1977), in which the NPS and its western and 
southern counter-spurs are interpreted as due to a dumbbell-shaped shock front
induced by a giant explosive event at the Galactic Center.
Such an impulsive energy input may have been
originated by a starburst 15 million years ago (Sofue 1984, 1994).
An alternative mechanism to cause such giant shells in the halo
would be a single energetic explosion at the Galactic nucleus (Oort 1977).
Giant shell structures could also be produced by a stellar-wind driven
coherent galactic wind (e.g., Heckman et al 1990). 
Since the cooling time in the halo is much longer than the shell's life time, 
any of these mechanisms will, however, result in a similar shocked shells in 
the halo, given the same amount of total energy.
Another alternative, traditional idea to explain the NPS is the local-shell 
hypothesis, in which the NPS is interpreted as due to a very unique
supernovae remnant of the largest diameter in the Galaxy 
(Berkhuijsen et al. 1971; Egger 1993; Egger and Aschenbach 1995; 
and the literature cited therein).  

The propagation of a shock front in the galactic halo can be simulated 
by applying a shock-tracing method developed by Sakashita (1971) and
M{\"o}llenhoff (1976) to a case 
of point explosion at a center of a disk surrounded by a halo and 
intergalactic uniform gas (Sofue 1984). 
The galactic-center explosion hypothesis  tries
to explain the NPS as well as the other spurs surrounding the galactic center
region by a single Galactic event, based not only on the morphology but also
on a distance estimate of NPS from soft-X ray extinction (Sofue 1994).

In the present paper, we revisit the galactic-center explosion hypothesis.
We will extend the arguments given in Sofue (1994), which were based on
the Winsconsin X-ray experiments data (McCammon et al 1983; 
McCammon and Sanders 1990).
We will discuss the origin of the NPS and related galactic spurs using
the 408-MHz all-sky radio data (Haslam et al 1982) and the ROSAT X-ray 
images (Snowden et al 1997). 
We simulate X-ray all-sky views based on the BHS (bipolar hyper-shell) model
in order to morphologically reproduce the ROSAT all-sky views at various
energy bands.
We also discuss the implication of the bipolar hyper shell in
dating and measuring the recent Galactic starburst.
We further propose to use the ROSAT all-sky data to probe the physics of
gas in the galactic halo and intergalactic space of the Local Group.

\section{All Sky Radio and X-ray Data}

\subsection{All-sky views}

In Fig. 1 we compare the radio and X-ray views of the whole sky in the
$(l,b)$ coordinates in Aitoff diagrams.
Fig. 1a shows an enhanced view of galactic radio spurs obtained 
by applying a relieving technique (Sofue 1993) to the 408 MHz all-sky 
map (Haslam et al. 1982). 
Fig. 1b is the ROSAT all-sky map in the R45 ($\sim$0.75 keV) 
band as reproduced from Snowden et al (1997). 
These figures demonstrates that the major galactic spurs both in radio and
X-rays are found in the central $100\Deg (\pm 50\Deg)$ region around
the Galactic Center. 
The North Polar Spur (NPS) and its western and southern counterparts 
compose giant $\Omega$ shapes, drawing  a dumbbell shape centered on 
the Galactic Center and necked at the galactic plane. 

--- Fig. 1 ---

\subsection{Radio spurs}

Fig. 2 shows a radio view in a 100\deg\ squared region around 
the Galactic Center. Here, a relieving method to enhance the 
spurs has been applied in the direction of longitude (left 
panel) and in the radial direction (right panel). 
The NPS comprises a well-defined radio arc anchored to the galactic plane at 
$l\sim \pm 20\Deg$, and draws a giant arc toward the North Galactic Pole.
The radio brightness along the NPS increases toward the galactic plane, 
attaining a maximum at $(l,b) \sim (20^\circ, 0^\circ)$.
The width of the spur (half-intensity length across) decreases toward the 
galactic plane, and therefore, the NPS ridge becomes sharper toward the 
galactic plane (Sofue and Reich 1979).  

--- Fig. 2 ---

The NPS draws a giant loop toward high latitudes, and returns to the 
galactic plane  where it merges with a spur emerging
from the galactic plane at $\l\sim 340\Deg$. 
We call this western spur NPS-West. 
A western half of Loop IV, which is highly asymmetric and lacks 
the eastern half, makes a part of NPS-West.
The NPS and NPS-West, thus, compose a giant $\Omega$ shape in the halo above 
the Galactic Center, with its axis roughly coinciding with the 
galactic rotation axis at $\l=0\Deg$.
A southern counterpart of the NPS is visible at $l \sim 20\Deg$, extending 
from $(l, b)\sim(20, 0)$ toward $(30, -30$, which we call the South Polar Spur 
(SPS). Also, a western counterpart of SPS is found at 
$(l, b) \sim (340, 0) $ to $(320, -30)$,  which we call SPS-West.

These four spurs, NPS, NPS-West, SPS, and SPS-West, 
are the most prominent features among the numerous galactic radio spurs. 
These four spurs compose a huge dumbbell-shape necked at the galactic plane
and are about symmetric with respect to the Galactic Center. 
We comment that Loop I, which has been defined as a complete loop of diameter 
120\deg centered on $(l, b)=(330\Deg, 30\Deg)$, is hard to trace in the 
present enhanced images (Fig. 1, 2).

\subsection{X-ray Spurs}

As shown in Fig. 1b, the R45-band (0.75 keV) X-ray intensity at $b>10\Deg$ has
a global enhancement around the Galactic Center, which is due to the 
high-temperature gas in the galactic bulge (McCammon et al 1983).
Snowden et al (1997) have further noticed cylindrical features in the
R45 and R67 (1.5 keV) band maps, which emerge from the central galactic disk
toward the halo.  They also attribute these features to high-temperature
gas around the galactic center.
This fact indicates that the local HI disk 
is transparent at $b> \sim 10\Deg$ for X-rays at $\ge 0.75$ keV. 

A giant shell structure of the NPS is clearly visible in the R45 and R67 bands.
In Fig. 3 we compare the radio and R45-band images of the NPS.
The western end of the X-ray NPS also returns to the galactic plane at 
$\l \sim 340\Deg$ (NPS-West), and draws a giant $\Omega$ together with the NPS.
Southern counterparts to these features are also visible.
Particularly, the SPS-West is clearly recognized in X-rays:
an R45-band spur  emerges from $l \sim 340\Deg$ toward the 
southern galactic pole, which is symmetric to the NPS about 
the Galactic Center. 
The SPS is also visible at $l\sim +30\Deg$, though fainter in R45-band, 
while it is more clearly visible in an R45/R67-ratio map. 
These northern and southern X-ray spurs are associated with the
radio spurs (NPS, NPS-West, SPS, and SPS-West), and draw a dumbbell-shape 
necked at the galactic plane. 
We may, hence, interpret that Snowden et al.'s "cylinder" 
would comprise the NPS, NPS-West, SPS and SPS-West.

--- Fig. 3 ---

The R12-band (0.25 keV) X-ray emission from the NPS is strongly absorbed below 
$b=60\Deg$, 
and is hardly visible below 30\deg\ (Snowden et al 1997).
The NPS shows up most clearly in the R45-band (0.75 keV),
while the emission is 
significantly absorbed in the galactic disk at $b<10\Deg$.
R67-band (1.5 keV) X-rays are also strongly absorbed near the galactic 
plane, indicating that the X-rays from the NPS originates in the 
space further than the HI disk (Sofue 1994).
Moreover, the X-rays become harder toward the galactic plane, as indicated by 
the clear decrease in the R45 to R67 intensity ratio toward the galactic plane.  
By comparing the R45 and R12-band intensities, we have shown that the HI mass 
toward $b=30\Deg$ to be  $7\times10^{20}$ H cm$^{-2}$, which is greater than 
the observed value ($5\times10^{20}$ H cm$^{-3}$).
Sofue (1994) has thus shown that the X-rays at $b \sim 30\Deg$ originate 
{\it beyond} the hydrogen layer, and the distance is greater than 0.6 kpc.
Namely, the NPS is an object which is located in the galactic halo.
The distance to the NPS has been subject to debates.
We now know that it is beyond 0.6 kpc, which would allow us two possible
interpretations:
One possibility is that the NPS is a local object at $\sim 1 $ kpc, originating
from supernova explosions at high altitude out of the galactic plane.
Another possibility is that it is a Galactic-scale object related to the
Galactic Center activity.

\section{Comparison of Bipolar Hyper-Shell Model with Data} 

\subsection{BHS Model}

We interpret the observed radio and X-ray features
in terms of the galactic-center explosion hypothesis.
Our idea is based on the symmetric appearance of the radio and X-ray shells 
with respect to the galactic plane and the Galactic Center, which compose a 
huge dumbbell shape apparently centered on the Galactic Center (GC).
We, here, assume that the center of the dumbbell coincides with the GC, and take
the distance to the GC to be 8 kpc.
Then, the radius  of each shell is several kpc. 
Our idea is also based on the fact that many spiral galaxies exhibit
galactic-scale outflows in the forms of dumbbell-shaped shocks,
bipolar cylinders, and galactic-scale jets, and we consider that the Milky Way
Galaxy would have experienced similar phenomena.

NGC 253 exhibits a dumbbell-shaped shells
in X-rays above and below the galactic plane, each about 5 kpc diameter 
(Vogler and Pietsch 1999a; Pietsch et al  1999) , which would be a result of 
a starburst and related outflow (Heckmann et al. 1990).
NGC 3079 exhibits a pair of radio continuum shells in the halo in
both sides of the nucleus, each about 3 kpc diameter, which is considered
to be an ejection from the central activity (Duric et al 1983).
It also exhibits an H$\alpha$ cone-shaped shell of radius about 
1 kpc, coaxial to the radio shells, most likely induced by an outburst from the 
nuclear region (Cecil 1999).
M82 is a starburst galaxy, which ejects a galactic-scale flow through
bipolar cylindrical jets (Nakai et al 1987).
We emphasize that all these out-of-plane features, including the hyper
shell in the Milky Way, require
a similar amount of total energy input in the central regions of
the galaxies, which is of the order of $10^{55}$ to $10^{56}$ ergs
(e.g. Pietsch et al 1999), or equivalent to $10^4$ to $10^5$ type II supernovae.

\subsection{Adiabatic-Shock Envelope Method}

The propagation of a shock wave through the galactic halo induced by a point 
energy injection at a galactic center can be calculated by applying the
shock envelope-tracing method of Sakashita (1971) and M{\"o}llenhoff.
They have extended the Laumbach and Probstein's (1969) method 
for tracing the evolution of a shock front to a case of axi-symmetric 
distributions of ambient gas.
The flow field is assumed to be locally radial, and the gas is adiabatic,
and, therefore, the heat transfer by radiation and counter pressure are neglected. 
The density contrast between the shock front and ambient gas is given
by $(\gamma+1)/(\gamma-1)=4$ for $\gamma=5/3$, where $\gamma$ is the adiabatic
exponent of the gas.
For a typical density of $10^{-3}$ H cm$^{-3}$ and temperature of
$10^7$ K for the shock-heated halo gas, the cooling time due to the
free-free thermal emission is approximately $5\times 10^8$ years.
Hence, the assumption of adiabatic gas is valid in our calculation,
in which the shell's life time is estimated to be 
of the order of $10^7$ years, as shown below.

The equation of motion of the shock wave is given by 
(M{\"o}llenhoff 1976):
$$ E=\int^R_0 {P \over{\gamma-1}} 4 \pi r^2 dr
+ \int^R_0 {1 \over 2} {\left( {\partial r \over \partial t} \right) ^2} 
\rho_0 4 \pi r_0^2 dr_0.
$$
Here, $E$ is the total energy of the explosion,
$P$ is the internal pressure, $\gamma$ is the adiabatic exponent, which 
is assumed to be 5/3 hereafter,
$\rho_0$ is the unperturbed ambient gas density, $r$ is the radius
from the explosion center, with suffix 0 denoting the quantities of the
unperturbed ambient gas, and $R$ is the radius of the shock front.
Assuming that the snow-plowed mass is strongly concentrated near the front,
the above equation leads to equation to express the shock radius $R$ as
follows (Sakashita 1971; M{\"o}llenhoff 1976):
\def\ga{\gamma}
\def\gm{(\gamma-1)}
\def\gp{(\gamma+1)} 
$$
E = {1 \over {3\gp^2}} \times
$$
$$
\left[
{{4(2\gamma-1)}\over \gm} J R {{\ddot R}} 
+ \left( 
 \left\{ 2 IR+ {{8\ga} \over\gp}+ 3 \right\} J 
+ {2 M \gp \over{\gm}}
\right)
{\dot R}^2
\right].
$$
Here, 
$$ I = \left[{4 \pi \over r_0} {{d \rho_0} \over {dr_0}} \right]_R, $$
$$ J = \int^R_0 \rho_0 4 \pi r_0^2 dr_0, $$
and
$$M=\rho_0 {{4 \pi} \over 3} R^3.$$

The unperturbed density distribution of gas is
assumed to  comprise a stratified disk, a halo with an exponentially 
decreasing density, and intergalactic gas with uniform density.
The hydrogen density in the galactic plane is taken 
to be 1 H cm$^{-3}$ in the solar vicinity.
The gas distribution is approximated by the following expression.
$$ \rho_0= \rho_1 {\rm exp}(-z/z_1) 
+ \rho_2 {\rm exp}(-z/z_2)
+ \rho_3.
$$
Here, suffices 1, 2 and 3 denote quantities for the disk, halo 
and intergalactic  gas, respectively, 
$\rho$ is the density, $z$ is the height from
the galactic plane, $z_i$ is the scale thickness of the disk and halo. 
Here, $\rho_1$, $\rho_2$, and $\rho_3$ are $\sim 1$, 0.01, and $10^{-5}$
 H cm$^{-1}$, and $z_1$ and $z_2$  are $\sim 0.1$ and 1 kpc, 
respectively.

This model has been applied to the Galactic Center in order to fit the 
North Polar Spur (Sofue 1984, 1994). 
The dumbbell-shaped shell structure and the NPS are well reproduced by a case 
in which the explosion energy is $E=3\times 10^{56}$ erg. 
Fig. 4 shows a calculated shock front at $t=10$ and 20 Myr as reproduced
from Sofue (1984). 
In this model, the expansion velocity of the shock front amounts to
several hundred \kms\, and the gas is heated up to $10^7$ K, emitting
soft X rays observable in the ROSAT energy bands at $\sim 0.25$, 
0.75 and 1.5 keV.

Numerical simulations of smaller-scale outflows from the galactic plane
have been obtained in hydrodynamic  scheme (e.g., Tomisaka and Ikeuchi 1986) 
and in MHD (magneto-hydrodynamic) treatment (e.g., Uchida et al 1985).
However, these simulations have been obtained only for smaller
scale outflows with scales less than one kpc, so that they cannot be applied to
the present case with a much larger-scale shells expanding from the upper halo 
to the intergalactic space.

--- Fig. 4 ---

\subsection{Radio and X-ray Simulations}

\subsubsection{Radio Sky}

We simulate radio and X-ray intensity distributions on the sky based on 
the bipolar hyper-shell model. 
In order to simulate the radio and X-ray emissivity in the
bipolar-hyper shells, as calculated above, we approximate
the shape of each half of the dumbbell-shaped shock by an ellipse.
Here, we calculate a case for a shell whose center is 
at $z=\pm6$ kpc and the radii 6 and 9 kpc 
in the radial and vertical ($z$) directions, respectively.  
The volume emissivity of synchrotron radiation is calculated from the
density contrast of the shocked gas, in which 
magnetic fields and cosmic-ray electrons are considered to be
compressed adiabatically with shock-compression of the halo gas,
and the volume emissivity of synchrotron radio radiation is assumed to be
proportional to $(\rho/\rho_0)^\beta$ with $\beta$ being approximately 4,
where $\rho$ and $\rho_0$ are
gas densities in the shocked shell and unperturbed halo gas, respectively. 
In the present simulation, 
the profile of radio emissivity perpendicular to the shell surface
is simply represented by an exponentially decreasing function behind the shock 
front toward the center with a scale thickness of 500 pc (Fig. 4). 
The emissivity also decreases with the height from the galactic
plane with a scale height of 3 kpc, corresponding to exponentially
decreasing density of the halo gas. 

In addition to the hyper shell, we assume the existence of 
a galactic disk of scale height of 0.5 kpc and scale radius of 6 kpc,
which is further embedded in a fatter disk with 3 and 8 kpc 
scale height and radius, respectively. The emissivity is, therefore, assumed
to have the form expressed by
$$ \epsilon =\epsilon_{\rm s}+ \epsilon_1+\epsilon_2
= \epsilon_{\rm s}+ \epsilon_0 \Sigma_{i=1}^2 {\rm exp}(-r/r_i - z/z_i),$$
where $\epsilon_{\rm s}$ is the emissivity in the shell as described above
(Fig. 4), $\epsilon_0$ is a constant, $\epsilon_1$ and $\epsilon_2$ 
represents the disk and fat components, respectively,  
$r_1=6$ kpc, $z_1=0.5$ kpc,  $r_2=8$ kpc, and $z_2=3$ kpc.
No extinction in the radio band is assumed throughout the galactic disk and halo.
In Fig. 5a we show the radio continuum result, and compare with the
observed 408 MHz all-sky map (Fig. 5c).
The global radio distribution is well reproduced by the model, and the
NPS is reproduced as a radio ridge emanating from the galactic disk.

--- Fig. 5 ---

\subsubsection{Radio Sky with Spiral Arms}

We further simulate a case in which the disk component comprises logarithmic
spiral arms, as illustrated in Fig. 6. 
The radio continuum emissivity is assumed to have a form
$$\epsilon_1
=\epsilon_0 {\rm exp}(-r/r_1-z/z_1) {\rm cos}(\theta-\eta log(r/r_0))^k.$$
Here, $\theta$ is the galacto-centric azimuthal angle, 
$\eta$ is the inverse of the pitch angle of spiral arms, 
$\eta=1/tan~p$ with $p=6^\circ$, $\alpha=5.1$ is a correction factor to fix the
total luminosity to be the same as that when no spiral arms are 
assumed, and $k=16$ is introduced to mimic a narrow condensation of 
the emissivity in the arms. 
The simulated result is shown in Fig. 5b.
In addition to the spurs due to the hyper shells as in Fig. 5a, 
there appear many spurs extending toward high latitudes in both sides of the
galactic plane.
These spurs are tangential views of bank-shaped spiral arms which are
extending into the halo. 
It is interesting to note that the tangential directions of the local spiral 
arms coincide with the western half of Loop I, 
Loop II and III, which was already pointed out earlier (Sofue 1976).
Simulated inner-arm spurs are also found to reproduce many
observed inner radio spurs, while their exact positional coincidence is
out of the present simplified spiral arm model.

--- Fig. 6 ---

\subsubsection{X-ray Sky}
 
X-ray emission from the hyper shells is assumed to be 
thermal free-free radiation, whose emissivity is given by 
$$\epsilon = n_{\rm e}^2 \Lambda 
\propto \rho^2 T^{1/2}.
$$
Here, $\Lambda$ is the cooling function of the gas.
Since the temperature is as high as $T\sim 10^7$ K in our simulation, 
the radiation is almost totally free-free, and the contribution by 
recombination lines from hydrogen, helium, and metals, which are 
significant at $T\sim 10^4-10^6$ K, are not significant.
For a typical density and temperature of about $10^{-3}$ H cm$^{-3}$
and $10^7$ K, respectively, the value of $\Lambda$ is $10^{-21}$ 
erg cm$^{-9}$ s$^{-1}$, and the emissivity is approximately
$10^{-27}$ erg cm$^{-3}$ s$^{-1}$.
A typical emission measure along the hyper shell ridge is
$n_{\rm e}^2 L \sim 10^{-2}$ cm$^{-6}$pc for a tangential pass in the shell
of about 2 kpc. 
For simplicity, we assume that the temperature of X-ray emitting gas
is  constant in the shocked shell. 
The assumption of constant temperature may not be a good approximation
for lower latitude regions, where the propagation velocity is
lower than that of the upper part, and, therefore, the temperature
is lower.
However, the low-latitude regions are strongly absorbed by the
HI gas layer, and such temperature gradient in the shell would not much
improve the accuracy of the present simulation.

In addition to the hyper shells, 
we assume a galactic ridge X-ray component of scale height of 500 pc, and
a bulge component of scale radius 1 kpc. 
Emission from ambient halo gas is neglected, because the temperature and density
will not be sufficiently high to emit X-rays in the present
energy bands.

The X-ray intensity, $I$, is calculated by a transfer equation given by
$$ dI= \epsilon_{\rm X} ds - \kappa I ds,$$
where the first term in the right-hand side is the emission measure with
$\epsilon_{\rm X}$ being the X-ray emissivity,
and the second term represents the absorption rate with
$\kappa$ being absorption coefficient, 
and $s$ is the distance along the line-of-sight.

The interstellar extinction of soft X rays occurs due to 
photoelectric absorption by metals, and has been
calculated for the solar metal abundance
by Morrison and McCammon (1983).
The energy dependence of the cross section per H atoms can be approximated by 
$\sigma \propto E^{-2.5}$ (Ryter 1996) in the present energy bands,
where $E$ is the photon energy of X rays.
The absorption coefficient $\kappa$ is, then, approximately represented by
$$\kappa = (n_{\rm H}/N_{\rm H}^0) (E/E_0)^{-2.5},  $$
where $n_{\rm H}$ is the number density of hydrogen atoms, 
and $N_{\rm H}^0$ is the $e$-folding column density of interstellar
neutral hydrogen at the photon energy $E_0$.
At $E_0=0.75$ keV, the cross section is 
$\sigma=4\times 10^{-22}$ cm$^2$ per H atom, which yields
an $e$-folding column density of $2.5\times10^{21}$ H atoms.
According to the transmission diagram of Snowden et al (1997),
the $e$-folding column density for R45 band X-rays ($E_0=0.75$ keV) 
is also read as 
$N_{\rm H}^0=3 \times 10^{21}$ H cm$^{-2}$. 
We adopt this value for the R45 band (0.75 keV). 
However, the ROSAT energy bands are rather broad, containing significantly 
higher and lower energies around the representative energies.
In our simulation, therefore, we adopt three representative cross sections
at R12, R45 and R67 bands, which are 10, 1 and 0.1 times the value at
0.75 keV, respectively.

The neutral hydrogen layer is assumed to have a density distribution 
expressed by
$$ n_{\rm H}=n_0{\rm exp} (-r/r_{\rm H} - z/z_{\rm H}),$$
where $n_0$ is the mean density in the solar vicinity, and
 is taken to be $n_0=1$ H cm$^{-3}$, $r_{\rm H}=8$ kpc is the scale radius
of the gas disk, and $z_{\rm H}=100$ pc is the scale height.

Fig. 7a shows the calculated intrinsic intensity distribution, where no 
galactic absorption is taken into account.
Fig. 7b is a result calculated for R45-band X-rays, which suffer 
from absorption by the galactic atomic hydrogen layer.
The observed absorption features near the galactic plane is well
reproduced by this model.
The hyper shells are observed as a set of double-horn features extending
toward the galactic poles, composing dumbbell-shaped ridges, and mimic
the observed X-ray spurs (NPS, NPS-West, SPS and SPS-West).
Fig. 7c shows a case in which the hyper shells are replaced by
bipolar parabolic cones, which are open toward the polar axes 
as illustrated in Fig. 4c.
The conical cylinder model appears, however, worse in order to reproduce 
the observed round shape of the North Polar Spur in radio and X-rays.
In both cases of hyper shells and conical cylinders, the observed cylindrical
spurs at lower latitudes in X-rays (Snowden et al 1997) are well reproduced.
The bulge component is also visible in the models, as the two enhanced regions
above and below the galactic center.

--- Fig. 7 ---

\subsubsection{X-ray Sky with Spiral Arms}

Fig. 8b shows results for a case in which the absorbing hydrogen gas 
is assumed to be condensed in logarithmic spiral arms of pitch 
angle 6$^\circ$ and the arm width is taken to be
1/5 of the arm separation, as illustrated in Fig. 6.
Hence, the peak hydrogen density in the local arm is about 5 H cm$^{-3}$.
The hydrogen density distribution is expressed by
$$ n_{\rm H}= \alpha n_0 {\rm exp}(-r/r_H -z/z_H) 
 {\rm cos}^k (\theta - \eta {\rm log}~r/r_0). $$
Here, $\theta$ and $\eta$ have the same meaning as above.   
Since the tangential directions of the spiral arms are asymmetric
around the galactic center, the absorption feature near to the galactic
plane is also asymmetric.

In order to examine how the X-ray sky looks like in different
energy bands, we have performed simulations for different values
of the absorption coefficient.
Fig. 8 shows the calculated results for three different $\kappa$ values:
(a) ten times the above value, (b) as above,  and  (c) one tenth.
Fig. 8a, b, and c, therefore, roughly correspond to X-ray views in the 
R12, R45, and R67 bands, respectively.
In Fig. 8, we compare the simulated results with the corresponding ROSAT images.

The R12-band model is characterized by a wide absorption lane along
the galactic plane, where X-rays from the galactic plane and bulge are almost
totally absorbed.
The hyper shells also suffer from strong absorption, while high-latitude
parts are still visible as the northern and southern "polar-caps", which
are indeed observed in the ROSAT R12-band image.
We also obtain a good reproduction of the observed ROSAT X-ray features in 
R45 and R67 bands, as discussed in the previous subsections.

\subsection{Summary of Comparisons Between Models and Data}

We have revisited the hyper-shell model of radio and X-ray spurs around the 
Galactic Center, and simulated the all-sky view of radio and X-ray emissions.
The model can mimic the observed all-sky views fairly well in radio (Haslam
et al 1982) and ROSAT X-rays (Snowden et al 1997), as shown in Fig. 5, 7 and 8.
The simulation could reproduce the following characteristic properties for
the galactic spurs :

{\parindent=0pt
(1) The major spurs in the whole sky are found around the Galactic Center 
at $330^\circ<l<30^\circ$ (NPS, NPS-West, SPS, and SPS-West). 

(2) These four spurs are located apparently symmetric with respect to the 
rotation axis of the Galaxy as well as to the galactic plane.

(3) Southern X-ray spurs exhibit cylindrical appearance.

(4) X-ray spurs in R45 (0.75 keV)- and R67 (1.5 keV) bands 
exhibit a sharp absorption lane along the galactic plane.

(5) R12-band (0.25 keV) spurs comprise northern and southern 
polar caps at high latitudes, $b>\sim 30-40^\circ$, 
while lower-latitude emission is hardly visible due to 
broad and strong absorption lane in the local galactic plane.

(6) An X-ray bulge is visible above and below the Galactic Center in
R45 and R67 bands.

(7) Besides the BHS-related spurs, a more number of radio continuum
spurs are observed in the tangential directions of the inner and local
spiral arms, with the most prominent spiral-arm spurs
emerging at $l \sim 80^\circ$ and $l\sim 260^\circ$ corresponding 
to the local arm.
}

\section{Discussion}

\subsection{Starburst Origin of Bipolar Hyper Shells}

\subsubsection{Time Scale and Energetics}

In our model, the energy injection at the Galactic Center has been assumed
to be impulsive, and the shock is strong enough to create a well-defined
shell structure. 
The time scale of the explosive event 
at the Galactic Center is assumed to be sufficiently shorter than
the expanding time scale of the hyper shell.
Since the expanding time scale is of the order of
$t\sim r/v \sim 10^7$ years for $r\sim$several kpc and $v\sim$300 \kms,
the explosion time scale should be shorter than a few million years.
The required total energy given to the interstellar gas is of the order
of $10^{55}-10^{56}$ ergs in order to heat the gas up to $\sim 10^7$ K
at the BHS front. 
Such impulsiveness and robustness of the energy release 
can be explained if the Galactic Center has experienced a starburst
about 15 million years ago, lasting for a few million years or shorter,
during which $\sim 10^5$ type II supernovae exploded. 
If this starburst model is correct, the ROSAT all-sky data can be used to
directly date and measure the energy of the recent starburst in our Galaxy.

\subsubsection{Hyper Shell's Neck in the Disk and the 2.4-kpc Expanding Ring}

By an MHD wave approximation of a blast wave from the Galactic Center
we have shown that a significant fraction of low-latitude front
focuses on the galactic plane, with the gas-flow paths being diffracted due to 
velocity gradient, which leads to an expanding gaseous ring in the disk 
(Sofue 1977).
The tangential directions of this expanding ring in the present model,
which coincide with the BHS dumbbell's neck, are at $l\sim \pm20^\circ$.
In fact, the root of the observed radio NPS crosses the galactic plane also 
at $l\sim \pm20^\circ$ (Sofue and Reich 1979).
It is interesting to point out 
that there is an expanding feature in the HI line emission
in coincidence with this tangential direction of the hyper shell,
known as the "3 kpc (2.4 kpc) expanding ring" of HI gas (Oort 1977).
Since the expanding velocity of the hyper shell in the disk region is
significantly decelerated, the ring will not not be heated up
to emit X rays, and, therefore, the ring may not be detectable in X-rays
at this low latitude. 

The Milky Way is also known to have another gaseous ring in the disk, the 
"4-kpc molecular ring", with high density
concentration of molecular and HI gases.
Although there is a suggestion about its origin as due to a resonance in
a barred potential (Nakai 1992),
it may also be possible to explain such a larger-radius ring
by focusing of an older expanding blast wave induced by
an earlier starburst recurrently occurring in the Galactic Center. 

\subsection{Alternative Interpretations}

\subsubsection{Alternative Impulsive Energy Injections at the Galactic Center}

We may consider some alternative mechanisms to cause an 
impulsive energy release in the Galactic Center region.
One possibility is a single gigantic explosion at the Galactic nucleus
(Oort 1977), possibly caused by an infall
of a star or its debris into the central massive black hole
(Genzel et al 1997; Ghez et al 1998). 
A stellar-wind driven Galactic wind (Heckman et al 1990) could also
produce a shell structure, as observed with ROSAT for
NGC 253 (Pietsch et al 1999). 
However, if the wind is steady and longer-lived than 15 million years,
the flow would have become already open-cone shaped, 
which does not appear to fit the observations.
Another possible mechanism is a "meteorite-like impact" by infalling
gas clouds from intergalactic space onto the galactic gas disk,
such as debris of the Magellanic Clouds after three-body dynamical 
interaction of the galaxies and the Clouds.
If several giant molecular clouds hit the gas disk at about the escaping velocity,
$\sim 400$ \kms, and the kinetic energy is converted into heat, the impact
will be equivalent to an explosion of energy of $\sim 10^{55}$ ergs.

Since the cooling time of the hyper shell, which is of the order of $10^9$ years, 
is much longer than the expansion time scale, any of these mechanisms will 
result in a similar shocked shells in the halo,
if the total input energy is approximately the same.

\subsubsection{On the Local-Shell Hypothesis}

Although we have proposed a unified model to explain the galactic 
spurs in terms of a single event at the Galactic Center,
we cannot exclude the possibility
that local objects such as nearby supernova remnants are 
superposed (Snowden et al 1997).
If we stand on the local-shell model, the sharp absorption lane along the 
galactic plane in R45 and R67 bands must be explained by an intrinsic
depression of intensity below $l\sim 10-20^\circ$.
We have in fact simulated intensity distributions in soft X-ray bands 
as expected from a local uniform shell of diameter 300 pc at 
a distance of 200 pc embedded in a hydrogen layer of a scale height 100 pc and 
local density 1 H cm$^{-3}$.
In R45 and R67 bands, no sharp absorption feature at $l<10-20^\circ$ was 
obtained by the simulation because of the negligible extinction for its proximity. 
The absorption feature can be reproduced, only if the shell
is not a unique emission source, but it is superposed by a more distant
emission such as from the Galactic bulge (Snowden et al 1997). 
In R12 band, the local-shell model could 
mimic the observation, similarly to the hyper-shell model,
because the distance does not affect the simulated results, since the 
absorption feature is produced very locally:
Regardless its distance, half the shell at high latitudes is visible in R12 band,
while the lower half at $b<\sim 30^\circ$ is absorbed by the local hydrogen layer.

A traditional idea to explain the NPS is the local supernova hypothesis,
which attributes the spur to a nearby shell of supernova remnant(s) of the 
largest diameter in the Galaxy, called Loop I (e.g., Berkhuijsen et al 1971; 
Egger and Aschenbach 1995).
The shell is the oldest supernovae remnant, 
and therefore, the probability to detect similar objects, 
if there exist any in the Galaxy, is much higher than that to find usual 
supernova remnants, almost two hundred of which are already discovered.  
Since the NPS is sufficiently bright in synchrotron radio emission,
with the typical brightness bing $\sim50$ K at 408 MHz, and is 
apparantely very large, such shells should have been easily detected, 
if it existed anywhere in the Galaxy. 
In this sense, the NPS is a very unique (peculiar) object, if we stand on the
local supernova hypothesis.
Note that Loops II and III, whose brightness temperatures are at most
about 4 K at 408 MHz, are an order of magnitude less luminous than NPS. 
Note also that the major parts of Loops II and III can be 
explained by local arm spurs in the present model (Fig. 5; see also Sofue 1976).

\subsection{Implications of the BHS Model as a Probe of Halo
and Intergalactic Gas Dynamics}

\subsubsection{Halo-Intergalactic Interface}

If the bipolar-hyper shell model applies, the all-sky ROSAT X-ray
and radio data can be used not only to measure the starburst, 
but also to probe the physics such as the density distribution and magnetic fields
in the galactic halo as well as those in the interface from the halo 
to intergalactic space.
The morphology of the hyper shell is sensitive to the density
distribution in the halo, and particularly, the shell shape in the
uppermost part manifests the halo-intergalactic density structure (Sofue 1984).
If the intergalactic gas density is very low, the shell will become
more open and conical, while if it is higher, the shell becomes
more compact and round.
The shell morphology, and therefore the density structure, 
is also dependent on the expanding velocity, which is
directly related to the gaseous temperature inferred from soft X-ray spectra.
More detailed hydrodynamical simulations with quantitative and morphological
fitting to the ROSAT data would provide us with much advanced knowledge about
the galactic halo and intergalactic space in the Local Group.

\subsubsection{Asymmetry of the Hyper Shell}

The observed NPS and its southern counter spurs are asymmetric in
morphology and intensities with respect to the galactic plane as well as to
the rotation axis of the Galaxy.
Such asymmetry can be attributed to anisotropy in the halo and intergalactic 
space.
Intergalactic wind, or equivalently, motion of the Galaxy in the Local Group,
may also cause an asymmetry of the hyper shell.
Hence, the morphological appearance of the hyper shell may be also used
to probe the anisotropy and winds in the halo and intergalactic space.

We have not taken into account the local
fluctuation of hydrogen gas and molecular clouds, which also cause 
asymmetry in the X-ray absorption appearance.
In fact the galactic layer is full of clouds, HI spurs, shells, and
holes (Hartmann and Burton 1997).
The Hydra spur of HI gas is a dense, leaned spur emanating from the galactic disk
at $l\sim 20^\circ$ toward $(l,b) \sim (0^\circ,30^\circ)$, crossing the
NPS root nearly perpendicularly, which would affect the absorption
feature of the NPS.
A more sophisticated modeling of the X-ray views and detailed
comparison with the ROSAT survey is subject for future simulations.
 
\subsection{Implications of the BHS Model for large-scale Structure in 
Nearby Galaxies}

As described in Section 3.1, many spiral galaxies exhibit similar
hyper shells in their halos (Sofue 1984): 
NGC 253' dumbbell-shaped shells
(Vogler and Pietsch 1999a; Pietsch et al  1999), 
NGC 3079's shells in radio continuum, H$\alpha$ and X-rays
(Duric et al 1983; Pietsch et al 1998; Cecil 1999).
M82's bipolar cylindrical jets (Nakai et al 1987).
NGC 4258's anomalous arms could also be due to some out-of-plane 
partial shells (van ALbada and van der Hulst 1982; Vogler and Pietsch 1999b).
We stress that these out-of-plane features, including our hyper
shell in the Milky Way, require a similar amount of total energy 
of the order of $10^{55-56}$, or $10^4$ to $10^5$ type II supernovae.
This fact suggest that the BHS starburst model would
be more general to model the extragalactic cases.
Simulations of ROSAT X-ray features in these objects based on
our model, which is in preparation, would provide us with 
a clue to date and measure the responsible explosive events at 
their centers, such as their starburst history as well as 
information about the gas dynamics in their halo-intergalactic interface.
Such simulations would also give a clue to compare the halo-intergalactic
gas physic and starbursts in external galaxies with those in our Milky Way.

\section{Summary}

We have revisited the BHS model of the North Polar Spur and related
galactic structures based on the 408-MHz radio continuum and 
ROSAT all-sky soft X-ray data.
We have shown that the NPS and its western and southern counter-spurs
draw a giant dumbbell-shape on the sky necked at the galactic plane.
The morphology and soft X-ray intensities
of the spurs can be interpreted as due to a shock front 
originating from a starburst at the Galactic Center
some 15 million years ago with a total energy of the 
order of $\sim 10^{56}$ ergs or $10^5$ type II supernovae. 

We have simulated radio continuum and soft X-ray skies based on the BHS 
galactic center starburst model.
Simulated all-sky distributions of radio continuum 
intensities  can well reproduce the radio NPS and related spurs, as well as
many other radio spurs in the tangential directions of spiral arms.
Simulated soft X-ray maps in 0.75 and 1.5 keV bands can reproduce the
ROSAT X-ray NPS, its western and southern counter-spurs, as well as the 
absorption layer along the galactic plane.
The observed R12 band polar-cap features are also well reproduced in our model.

If the present BHS model for the spurs is correct, we may be able
to use the ROSAT all-sky maps to probe the gas dynamics in the halo-intergalactic
interface.
We may also be able to directly date and measure the energy of a recent
Galactic Center starburst in the Milky Way, which can also be compared
with similar BHS phenomena in nearby starburst galaxies.

\vskip 5mm

Acknowledgement: The author would like to thank the ROSAT experiment staff
for making the FITS formatted data of the X-ray survey available via an
internet link to "SkyView" operated by NASA.
The author is indebted to the anonymous referee for the
invaluable comments and suggestions, with which the paper has been
very much improved. 

\section*{References}

\r Berkhuijsen, E., Haslam, C. and Salter, C. 1971 AA 14, 252.

\r Cecil, G. 1999 NAOA Preprint No. 850 (in "Imaging the Universe in Three
Dimensions", eds. W. van Breugel and J. Bland-Hauthorn, ASP Conf. Series).

\r  Duric, N.,  Seaquist, E.R.,  Crane, P.C.,  Bignell, R.C.,  
Davis, L.E.,  1983,  Ap.J.L,   273,  L11.

\r Egger, R. J., and Aschenbach, B. 1995 AA 294, L25.

\r Egger, R. 1993, Ph. D. Thesis, Univ. Munich, MPE Report No. 249.
 
\r Genzel, R., Eckart, A., Ott, T., and Eisenhauer, F.
	1997, MNRAS 291, 219.

\r Ghez, A., Morris, M., Klein, B. L., Becklin,E.E. 
	1998, ApJ 509, 678.

\r  Haslam, C.G.T.,  Salter, C.J.,  Stoffel, H.,  Wilson, W.E.,  1982,  AA 
Suppl. 47,  1.

\r Hartmann, D. and Burton, W.B. 1997, in Atlas of galactic neutral hydrogen,
Cambridge University Press, Cambridge.

\r Heckman, T. M., Armus, L., Miley, G. K., 1990, ApJS 74,  833. 

\r Laumbach, D. D., and Probstein, R. F. 1969 J. Fluid Mecha. 35, 53.

\r  McCammon, D. and Sanders,  1990, ARAA 28, 657.

\r  McCammon, D., Burrows, D.N., Sanders.W. T., and Kraushaar, W. L. 1983, 
Ap.J. 269, 107.

\r M{\"o}llenhoff, C. 1976 AA 50, 105.

\r Morrison, R. and McCammon D. 1983 ApJ 270, 119.
 
\r Nakai, N. 1992, PASJ 44, L27.

\r Nakai, N., M.Hayashi, Handa, T., Sofue, Y., Hasegawa, T., and Sasaki, M.
 PASJ 39, 685, 1987.

\r Oort, J.H. 1977 ARAA 15, 259.

\r Pietsch, W., Trinchieri, G., and Vogler, A 1998 AA 340, 351.

\r Pietsch, W., Vogler, A., Klein, U. and Zinnecker, H.   1999 AA in press.

\r Ryter, Ch. E. 1996 ApSpSc 236, 285.

\r Sakashita, S. 1971 ApSpSc 14, 431.

\r  Snowden, S. L., Egger, R., Freyberg, M. J., McCammon, D.,Plucinsky, P. P.,
  Sanders, W. T., Schmitt, J. H. M. M., Tr\"umpler, J., Voges, W. H.  
1997 ApJ. 485, 125 
 
\r Sofue, Y. 1976, AA 48, 1. 

\r  Sofue, Y. 1977, AA 60, 327. %MHD NPS

\r  Sofue, Y. 1984,  PASJ  36,  539. 
 
\r Sofue, Y.  1993, PASP, 105, 308. 

\r Sofue, Y.  1994, ApJ.L., 431, L91 

\r Sofue, Y. and Reich, W. 1979 AAS 38, 251

\r Tomisaka, K. and Ikeuchi, S. 1988 ApJ 330 695

\r Uchida, Y., Shibata, K., and Sofue, Y. 1985  Nature 317, 699

\r van Albada, G.D. and van der Hulst, J.M., 1982 AA 115, 263.

\r Vogler, A., and Pietsch, W., 1999a AA 342, 101. %Rosat N253 I

\r Vogler, A., and Pietsch, W., 1999b AA in press.

\newpage

\parindent=0pt
\parskip=4mm
Fig. Captions 

Fig. 1: (a) Top: Enhanced view of galactic spurs in the 408-MHz radio continuum,
as obtained by applying a relieving method to the Bonn-Parkes all-sky 
survey in the galactic coordinates (Haslam et al 1982). 
Major features discussed in the text are indicated by arrows.
The Galactic Center is at the map center, and the longitude increases
toward the left with the both side edges being at $l=180^\circ$.
The Galactic North Pole is to the top, and South Pole at the bottom. 

 (b) Middle: An X-ray all-sky image in the R45 (0.75 keV) band as reproduced 
from the ROSAT survey (Snowden et al 1997). 

(c) Bottom: Schematic overlay of the bipolar hyper shell (BHS) 
on the ROSAT 0.75 keV map.

Fig. 2: 408 MHz radio map in a $\pm 50\circ$ square region around 
the Galactic Center. 
A relieving method has been applied 
in the direction of longitude (left panel) and radial direction (right panel). 
Symmetric sets of spurs emerging from $l\sim 20\Deg$ and $\sim 340\Deg$ are 
recognized, which labeled as NPS, NPS-West, SPS and SPS-West (see the text). 
  
Fig. 3: Enhanced radio (left) and X-ray (right) images of the North Polar Spur.
Displayed area is $\pm 50^\circ$ square centered at $(l,b)=(0,30^\circ)$.

Fig. 4: (a) Calculated shock front in the galactic halo at 1 and 
$2\times10^7$ yr after an explosion and/or a starburst at the nucleus with a 
total energy of $3 \times 10^{56}$ erg (Sofue 1984).  

(b) Same as (a) but showing the shock front every 2 Myr.

(c) Schematic view of a bipolar conical shock front for a
case of lower gas density in the upper halo and intergalactic space.

Fig. 5: Simulated all-sky radio intensity distribution in the 
galactic coordinates according to the bipolar hyper shell model. 
Intensity scales are relative, and absolute values are arbitrary. 

(a) Hyper shells with a flat galactic disk and a thick disk without spiral arms.

(b) Hyper shells with a galactic disk comprising logarithmic spiral arms 
and a thick disk.

(c) Observed 408 MHz all-sky map in gray scale (Haslam et al 1982). 
Thick arrows indicate the bipolar hyper shell. 
Thin arrows indicate the tangential directions of the local and inner 
spiral arms, where radio spurs are emerging from the disk toward the halo.

Fig. 6: Logarithmic spiral arms with pitch angle of 6$^\circ$ and condensation
of about 5 times the averaged value.

Fig. 7: Simulated all-sky distributions in the R45 (0.75 keV) X-ray bands.

(a) Intrinsic intensities of the hyper shells and disk 
without interstellar extinction.

(b) Intensity distribution for an exponential
hydrogen absorbing layer, but without spiral arms. 

(c) Same as (b) but for a parabolic conical cylinders. 
The $\Omega$ shape of the North Polar Spur is better reproduced
by the hyper shell model of (b)

Fig. 8: Simulated all-sky X-ray views in R12, R45
and R67 bands (left) compared with the corresponding ROSAT sky views (right).
The absorbing hydrogen layer is assumed to comprise logarithmic spiral
arms as in Fig. 6.
The polar caps in 0.25 keV band, dumbbell-shaped morphology in 0.75  and 1.5 keV
bands with the absorption features along the galactic plane are well reproduced
by the BHS model calculations

\end{document}